\documentclass[twocolumn,aps,showpacs]{revtex4-1}
\usepackage[latin9]{luainputenc}
\usepackage{amsmath}
\usepackage{graphicx}
\usepackage{esint}
\usepackage{hyperref}
\usepackage{color}
\definecolor{med-blue}{RGB}{25,25,112}
\hypersetup{colorlinks, linkcolor={red},citecolor={blue}, urlcolor={blue}}

\makeatletter
\@ifundefined{textcolor}{}
{%
 \definecolor{BLACK}{gray}{0}
 \definecolor{WHITE}{gray}{1}
 \definecolor{RED}{rgb}{1,0,0}
 \definecolor{GREEN}{rgb}{0,1,0}
 \definecolor{BLUE}{rgb}{0,0,1}
 \definecolor{CYAN}{cmyk}{1,0,0,0}
 \definecolor{MAGENTA}{cmyk}{0,1,0,0}
 \definecolor{YELLOW}{cmyk}{0,0,1,0}
}

\usepackage{color}

\newcommand{\be}{\begin{equation}}
\newcommand{\ee}{  \end{equation}}
\newcommand{\ba}{\begin{eqnarray}}
\newcommand{\ea}{  \end{eqnarray}}
\newcommand{\ket}[1]{\left|#1\right>}
\newcommand{\bra}[1]{\left< #1 \right|}

\makeatother

\begin{document}

\title{Nonlinear dynamics of a two-level system of a single spin driven
beyond the rotating-wave approximation}

\author{K. Rama Koteswara Rao and Dieter Suter}

\affiliation{Fakult{ä}t Physik, Technische Universit{ä}t Dortmund, D-44221
Dortmund, Germany}

\date{\today}

\pacs{03.67.Lx, 76.70.Hb, 33.35.+r, 61.72.J-}
\begin{abstract}
Quantum systems driven by strong oscillating fields are the source
of many interesting physical phenomena. In this work, we experimentally
study the dynamics of a two-level system of a single spin driven in
the strong-driving regime where the rotating-wave approximation is
not valid. This two-level system is a subsystem of a single Nitrogen-Vacancy
center in diamond coupled to a first-shell $^{13}$C nuclear spin
at a level anti-crossing point. This near-degeneracy occurs in the
$m_{s}=\pm1$ manifold of the electron spin when the energy level
splitting between the $m_{s}$ = $-1$ and $+1$ states due to the
static magnetic field is $\approx$ 127 MHz and thus equal to the splitting
due to the $^{13}$C hyperfine interaction. The transition frequency
of this electron spin two-level system in a static magnetic field
of 28.9 G is 1.7 MHz and it can be driven only by the component of
the RF field along the NV symmetry axis. Electron spin Rabi frequencies
in this system can reach tens of MHz even for moderate RF powers.
The simple sinusoidal Rabi oscillations that occur when the amplitude
of the driving field is small compared to the transition frequency
evolve into complex patterns when the driving field amplitude is comparable
to or greater than the energy level splitting. We observe that the
system oscillates faster than the amplitude of the driving field and
the response of the system shows multiple frequencies. 
\end{abstract}

\keywords{NV center, Strong driving, Rotating-wave approximation}
\maketitle

\section{Introduction}

Two-level systems are the basis for many important fundamental concepts
in diverse areas of physics, including various types of resonance
phenomena \cite{Slichter,AE}. A quantum bit or qubit, which is realized
physically by a two level quantum system, is the fundamental building
block of quantum computation and quantum information protocols \cite{NC}.

A two-level system is equivalent to a spin-1/2 particle precessing
in a static magnetic field, where the spin precession frequency is
equal to the frequency separation between the two levels. Resonance
occurs when an oscillating field with a frequency equal to the spin
precession frequency is applied in a direction perpendicular to the
static field. In most experimental situations, the driving field is
a linearly oscillating one and it can be written as a sum of a co-
and a counter-rotating field with respect to the spin precession.
Most of the fundamental concepts in optical and magnetic resonance
are derived in the so-called weak-driving regime, where the amplitude
of the driving field is much smaller than the spin precession frequency.
In this regime, the effect of the counter-rotating field component can
be neglected, which is known as the rotating-wave approximation \cite{AE}.
The co-rotating field component causes harmonic oscillations between
the two levels, which are known as Rabi oscillations. In the strong-driving
regime, where the amplitude of the driving field is of the order or
greater than the spin precession frequency, the rotating-wave approximation
breaks down and the counter-rotating field component, which can not
be neglected any more, leads to many interesting physical phenomena
\cite{CDT,BoydTLS,NoriCDT,Chil2010PRA}. In this regime, the system's
dynamics are highly anharmonic and nonlinear, but not chaotic \cite{TLSchaosPomeau,TLSchaosEidson}.
Recently, the strong-driving regime has been of particular interest
to quantum information processing because of the ultra-fast quantum
gates possible in this regime \cite{Fuchs2009,BRWAJelezkoPRA,BRWAJelezko,BRWASCqubit,MechDriveNV,BRWASi}.

In this work, we experimentally observe the anharmonic and non-linear
dynamics of a two-level system of a single solid-state spin driven
in the strong-driving regime. The system of interest is a single Nitrogen-Vacancy
(NV) center (electron spin $S$=1) in diamond coupled to a first-shell
$^{13}$C nuclear spin ($I$=1/2) and the intrinsic $^{14}$N nuclear
spin ($I$=1) of the center, which together form an 18-level system.
A small static magnetic field is applied at an orientation such that
the energy level splitting due to the Zeeman interaction of the electron
spin is equal to the corresponding splitting due to the $^{13}$C
hyperfine interaction, which is approximately 127 MHz. At this point,
Level Anti-Crossings (LACs) occur between the $m_{s}=-1$ and $+1$
spin-sublevels due to the strong non-secular components of the Hamiltonian.
Far away from the LAC, electron spin transitions between these spin-sublevels
are forbidden. At the LAC point, the $m_{s}=+1$ and $-1$ spin states
are completely mixed and a transition can be excited between these
levels by the component of an applied radio-frequency (RF) field that
is parallel to the NV symmetry axis ($z$-axis). The transition frequency
of this electron spin transition is 1.7 MHz in a static magnetic field
of 28.9 G. In this system, the electron spin Rabi frequencies can
reach tens of MHz even for moderate RF powers. Hence, this two-level
system forms a very interesting platform for studying dynamics in
the strong-driving regime. In the weak-driving regime, we observe
clear sinusoidal Rabi oscillations of this two-level system. In the
strong-driving regime, the dynamics become complex as the system oscillates
with multiple frequencies and faster than the amplitude of the applied
RF field.

All the experiments of this work have been performed using a home-built
confocal microscope for optical excitation and detection of single
NV centers and RF and microwave (MW) electronics for resonant excitation.
The diamond crystal used has natural abundance of $^{13}$C atoms
and a nitrogen impurity concentration of $<$ 5 ppb.

\section{System and Hamiltonian}

\begin{figure}[t]
\centering \includegraphics[width=8.5cm]{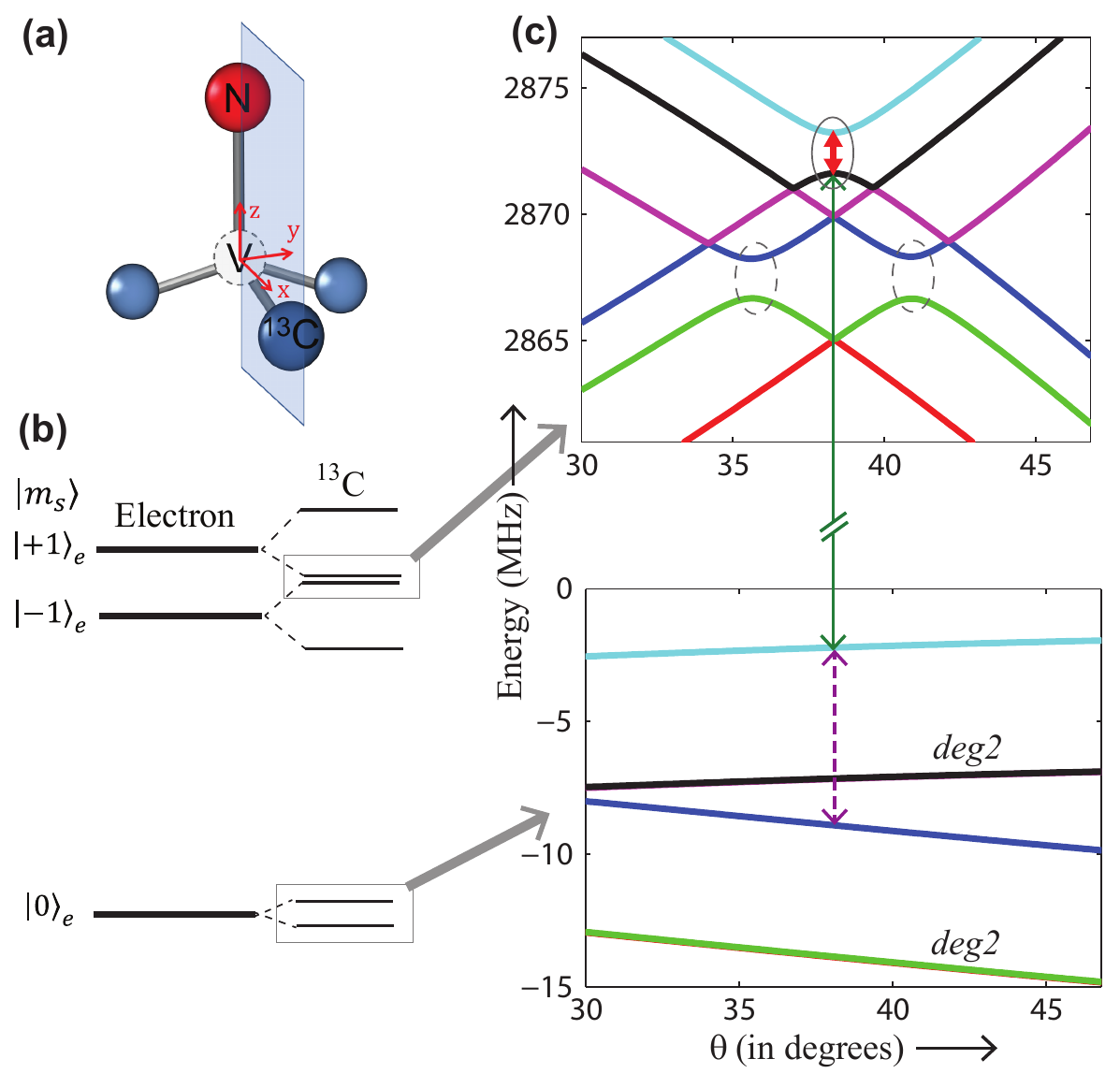} \caption{(a) Structure of an NV center coupled to a first-shell $^{13}$C nuclear
spin. (b) Simplified energy level diagram of this system considering
only the electron and the$^{13}$C nuclear spins. (c) The energy levels
marked by gray rectangles in (b) are plotted as functions of $\theta$
of a magnetic field of strength, $B$=28.9 G, and $\phi$=0$^{\circ}$.
Here, interactions due to the $^{14}$N nuclear spin are also included.
The energy levels labeled by \textit{deg2} are nearly doubly degenerate.
The transitions marked by thick red and thin green arrows are electron
spin transitions and the one marked by dashed violet arrow is a $^{13}$C
nuclear spin transition.}
\label{Englvl} 
\end{figure}

The structure of an NV center coupled to a first-shell $^{13}$C
nuclear spin is illustrated in Fig. \ref{Englvl}(a) along with it's
coordinate system, which is defined as follows: The NV symmetry axis
is defined as the $z$-axis, the $x$-axis is perpendicular to this
axis and lies in the plane containing the nitrogen, the vacancy and
the $^{13}$C atom, and the $y$-axis is perpendicular to this plane.
The total Hamiltonian of the system including the intrinsic $^{14}$N
nuclear spin of the center in this coordinate system can be written
as

\begin{align}
{\cal H}= & DS_{z}^{2}+\gamma_{e}\mathbf{B}\cdot\mathbf{S}+\gamma_{n1}\mathbf{B}\cdot\mathbf{I_{1}}+\gamma_{n2}\mathbf{B}\cdot\mathbf{I_{2}}\nonumber \\
 & +PI_{2z}^{2}+\mathbf{S}\cdot\mathcal{A}_{1}\cdot\mathbf{I_{1}}+\mathbf{S}\cdot\mathcal{A}_{2}\cdot\mathbf{I_{2}}.\label{eq:Hamiltonian}
\end{align}

Here, $\mathbf{S}$, $\mathbf{I}_{1}$, and $\mathbf{I}_{2}$ represent
spin angular momenta of the electron, $^{13}$C and $^{14}$N nuclei
respectively, and $\gamma_{e}$, $\gamma_{n1}$, and $\gamma_{n2}$
the corresponding gyromagnetic ratios. $D=2870.2$ MHz is the zero-field
splitting, and ${\mathbf{B}}=B(\sin\theta\cos\phi,\ \sin\theta\sin\phi,\ \cos\theta)$
is the magnetic field vector, $\theta$ and $\phi$ represent its
polar and azimuthal angles. $P=-4.95$ MHz \cite{Bajaj14Nquadrapole}
is the quadrupolar splitting of the $^{14}$N nucleus, and $\mathcal{A}_{1}$
and $\mathcal{A}_{2}$ represent hyperfine tensors of the $^{13}$C
and $^{14}$N nuclear spins respectively with the electron spin. The
values for the hyperfine tensor components have been taken from previous
work. For the $^{13}$C nuclear spin, these are $\mathcal{A}_{1zz}=128.9$,
$\mathcal{A}_{1yy}=128.4$, $\mathcal{A}_{1xx}=189.3$, and $\mathcal{A}_{1xz}=24.1$
MHz \cite{Koti2016PRB} and for the $^{14}$N nuclear spin these are
$\mathcal{A}_{2zz}=-2.3$ MHz and $\sqrt{\mathcal{A}_{2xx}^{2}+\mathcal{A}_{2yy}^{2}}=-2.6$
MHz \cite{Mansion14Nhyperfine,Felton2009,PaolaPRB2015}. All the other
components of the hyperfine tensors are zero due to the symmetry of
the system.

Now, we analyze the LACs that occur between the $m_{s}=-1$ and $+1$
spin states due to the strong non-secular components of the Hamiltonian
when the energy level splitting between them due to the Zeeman interaction
of the electron spin is roughly equal to the energy splitting due
to the $^{13}$C hyperfine interaction: 2$\gamma_{e}B\cos\theta$
is $\approx$ 127 MHz. Fig. \ref{Englvl}(b) shows a simplified energy
level diagram, considering only the electron and $^{13}$C nuclear
spins. In this diagram, the energy levels relevant for the present
work are marked with gray rectangles. These specific energy levels,
after including the interaction with the $^{14}$N nuclear spin, have
been plotted as a function of $\theta$ of a static magnetic field
of strength, $B$=28.9 G, and $\phi$=0$^{\circ}$ in Fig. \ref{Englvl}(c).
For this magnetic field strength, the LAC condition ($2\gamma_{e}B\cos\theta\approx127$
MHz) corresponds to $\theta=38.4^{\circ}.$ As can be seen from Fig.
\ref{Englvl}(c), there are several LAC points near $\theta$=38.4$^{\circ}$.
Out of these, the LACs relevant for the present work have been marked
with gray ovals. These three LACs differ mainly with respect to the
state of the $^{14}$N nuclear spin. Here, we discuss specifically
the LAC at $\theta$=38.4$^{\circ}$, which is marked with the solid
gray oval and corresponds to $m_{I_{2}}=0$. \textit{The two energy
levels shown in this gray oval represent the two-level system of which
we study the strong-driving dynamics}. Away from the LAC, the electron
spin states of these two levels are $m_{s}=+1$ and $-1$, and the
transition between them is forbidden. At the LAC point, there is a
strong mixing between the electron spin states and the energy eigenstates
become 
\[
\ket{\psi_{1,2}}\approx\ket{\frac{\ket{-1}\pm\ket{+1}}{\sqrt{2}},-\frac{1}{2},0},
\]
 where $-\frac{1}{2}$ and $0$ represent spin states of the $^{13}$C
and $^{14}$N nuclei respectively. Consequently, a transition can
be excited between these two levels by the $z$-component (parallel
to the NV axis) of the RF field as $|\bra{\psi_{1}}S_{z}\ket{\psi_{2}}|\approx1$.

The Hamiltonian of this two-level system can be written as 
\begin{equation}
{\cal H}_{\textrm{TLS}}(t)=\frac{1}{2}\omega_{0}\sigma_{x}+\omega_{1}\cos(2\pi\omega t)\sigma_{z},\label{eq:TLS}
\end{equation}
where, $\omega_{0}=1.7$ MHz is the transition frequency, and $\omega_{1}$
and $\omega$ are the amplitude and frequency of the driving field
respectively. $\sigma_{x}$ and $\sigma_{z}$ are the Pauli spin matrices.
In this system, $\omega_{1}$ can have values that are more than an
order of magnitude larger than $\omega_{0}$ even for moderate RF
powers and hence it forms a very interesting system for experimental
studies of strong-driving dynamics.

As discussed in Section I, the linearly oscillating driving field,
$\cos(2\pi\omega t)\sigma_{z}$ can be written as a sum of co- ($\frac{1}{2}\cos(2\pi\omega t)\sigma_{z}-\frac{1}{2}\sin(2\pi\omega t)\sigma_{y}$)
and counter-rotating ($\frac{1}{2}\cos(2\pi\omega t)\sigma_{z}+\frac{1}{2}\sin(2\pi\omega t)\sigma_{y}$)
field components. In the weak driving regime i.e., $\omega_{1}\ll\omega_{0}$,
the rotating-wave approximation is valid and the counter-rotating
field component can be neglected while the co-rotating field component
drives Rabi oscillations of the system. These oscillations, damped
by relaxation processes, were first observed in nuclear spins by Torrey
\cite{Torrey1949}. The neglected counter-rotating field component
was shown to shift the resonance frequency of the system by $\omega_{1}^{2}/4\omega_{0}$,
which is known as Bloch-Siegert shift \cite{B-Sshift1940,AutlerTownes}.
This shift, however, is very small when $\omega_{1}\ll\omega_{0}$,
and when $\omega_{1}$ is of the order or greater than $\omega_{0}$
(strong-driving regime), the approximations that are used to derive
the shift themselves are not valid. In this strong-driving regime,
the rotating-wave approximation breaks down and both the co- and counter-rotating
field components drive the system dynamics in a non-trivial way. 

\begin{figure}
\includegraphics[width=8cm]{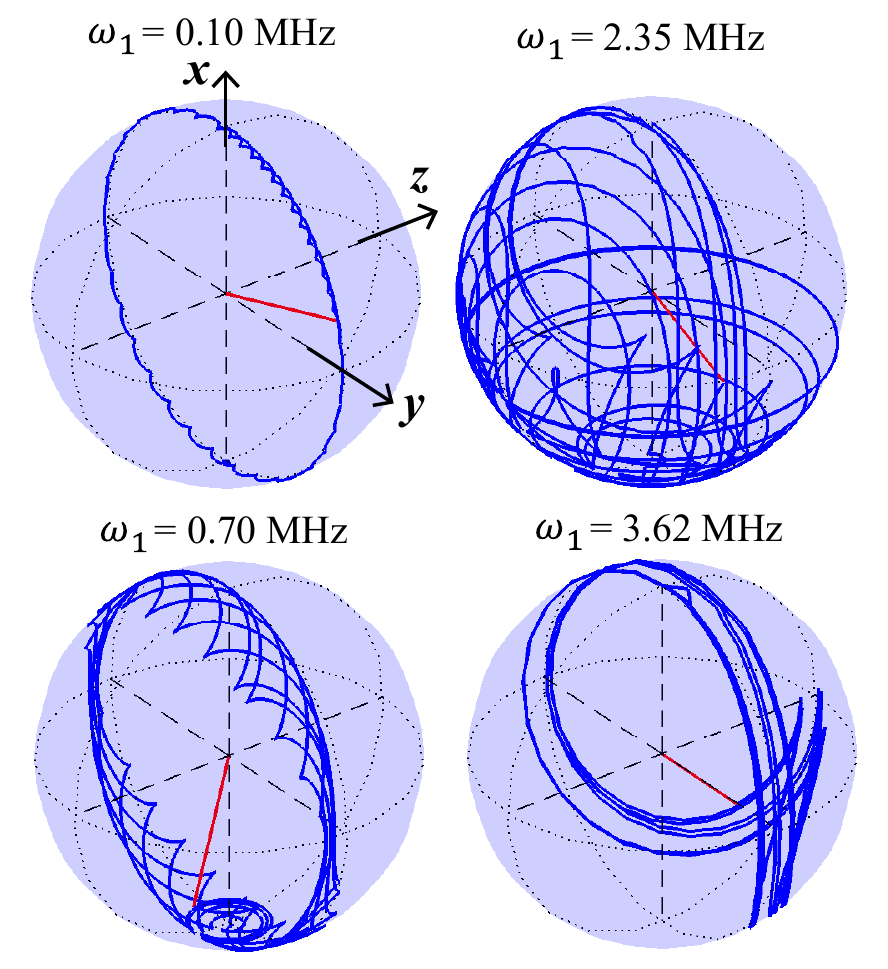}\caption{Bloch sphere representation of the simulated on-resonance ($\omega=\omega_{0}=1.7$
MHz) rotating-frame dynamics of the system of Eq. \ref{eq:TLS} for
different values of $\omega_{1}$.}

\label{BSrot}
\end{figure}

The Bloch sphere representation of the simulated on-resonance ($\omega=\omega_{0}$)
dynamics of the above two-level system for different values of $\omega_{1}$,
in a frame rotating with the same frequency ($\omega$) and direction
of the co-rotating field component, are shown in Fig. \ref{BSrot}.
The system is initialized into the $\left|+x\right\rangle $ eigenstate
of $\sigma_{x}$ . When $\omega_{1}=0.10$ MHz ($\ll\omega_{0}=1.7$
MHz), where the rotating-wave approximation is valid, the system rotates
on the surface of the sphere in the $xy$-plane with a frequency $\omega_{1}$.
When $\omega_{1}=0.70$ MHz (comparable to $\omega_{0}$), the system's
dynamics deviate from the $xy$-plane due to the counter-rotating
field component. When $\omega_{1}=2.35$ MHz ($>\omega_{0}$), where
the rotating-wave approximation is no longer valid, the spin trajectory
occupies a large fraction of the surface of the sphere and when $\omega_{1}=3.62$
MHz, the system doesn't flip to the $\left|-x\right\rangle $ state
completely for any finite time $t$.

\section{Experiments}

\begin{figure}
\includegraphics[width=8cm]{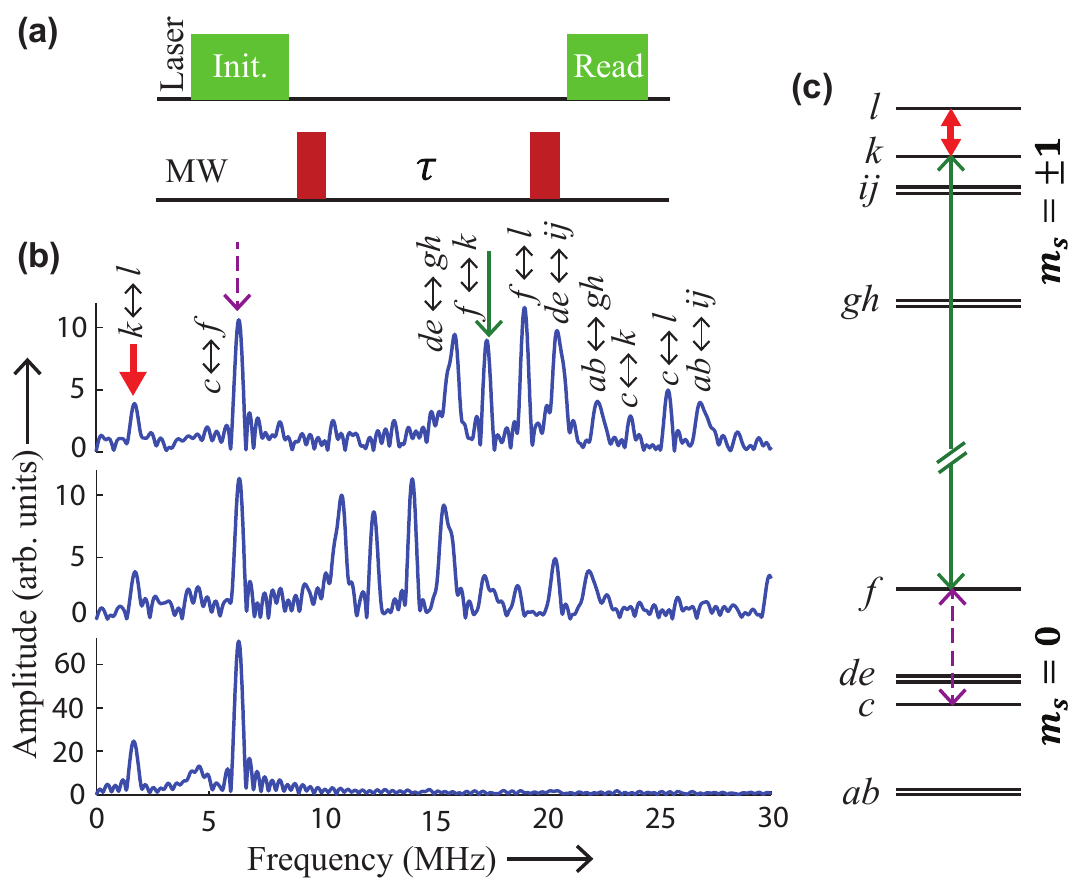}

\caption{(a) Pulse sequence used to measure Ramsey fringes. (b) Fourier transforms
of Ramsey fringes, recorded at the LAC point $\theta$=38.4$^{\circ}$
and $\phi$=0$^{\circ}$. For the top and middle traces, the flip
angle of the MW pulses was $\pi/2$ and the frequency detuning ($\nu_{d}$)
was 20 and 15 MHz respectively. For the bottom trace, the flip angle
was $\pi$. The spectral lines marked by thick red, thin green, and
dotted violet arrows represent electron spin transitions of frequency
1.7 and 2873.9 MHz, and a $^{13}$C nuclear spin transition of frequency
6.4 MHz respectively. (c) Corresponding energy level diagram at the
LAC point. Energy levels labeled by two letters (e.g., \textit{ab})
are nearly doubly degenerate.}

\label{Spec}
\end{figure}

An experimental study of this two-level system requires precise orientation
of the magnetic field with respect to the NV coordinate system. In
this work, this was achieved by a permanent magnet placed at a fixed
distance from the NV center and rotated around two orthogonal axes
crossing at the NV center \cite{Koti2016PRB}. The magnetic field
$B$ was 28.9 G, oriented at $\theta=38.4^{\circ}$ and $\phi=0^{\circ}$,
which corresponds to a LAC point (Fig. \ref{Englvl}(c)). The experimental
electron spin resonance (ESR) spectra between the energy levels marked
by gray rectangles in Fig. \ref{Englvl}(b) at this magnetic field
orientation are shown in Fig. \ref{Spec}(b). These were recorded
as Fourier transforms of Ramsey fringes by using the pulse sequence
shown in Fig. \ref{Spec}(a). All the spectral lines are labeled in
comparison with the corresponding energy level diagram illustrated
in Fig. \ref{Spec}(c). The frequency of the MW pulses was 2876.6
MHz. 
The phase of the second MW pulse (-2$\pi\nu_{d}\tau$) was varied
with respect to that of the first pulse as a linear function of $\tau$
(delay between the pulses) such that an artificial detuning ($\nu_{d}$)
is introduced in the spectra. The top trace of these spectra was recorded
with a frequency detuning ($\nu_{d}$) of 20 MHz. In this spectrum,
along with the electron spin transitions between the $m_{s}=0$ and
$m_{s}=\pm1$ spin states (transitions in the range 15-30 MHz), two
other transitions also appear. Of these, the transition that appears
at 1.7 MHz (marked by the thick red arrow) is an electron spin transition
between two levels of the $m_{s}=\pm1$ manifolds. These two levels,
which are shown inside the solid gray oval in Fig. \ref{Englvl}(c),
are the levels that we use to study the strong-driving dynamics. The
other transition, which appears at 6.4 MHz (marked by the violet dashed
arrow) is a $^{13}$C nuclear spin transition of the $m_{s}=0$ state.
\begin{figure}[b]
\centering \includegraphics[width=7cm]{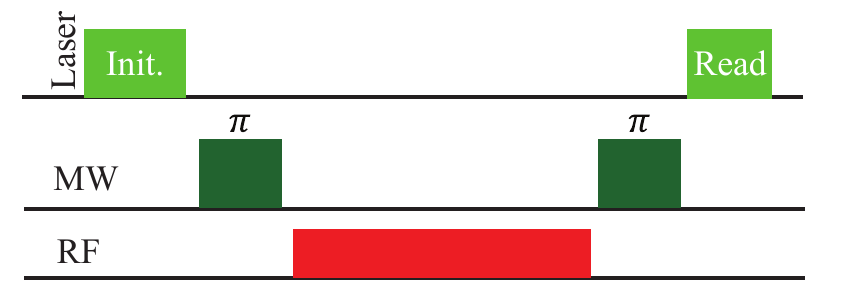} \caption{Pulse sequence for observing the dynamics of the two-level system.}
\label{pulseq} 
\end{figure}

These two transitions  appear in the ESR spectrum indirectly as zero-quantum
transitions \cite{Koti2016PRB}. For example, the transition between
the two levels of the $m_{s}=\pm1$ manifold can be excited by simultaneously
driving the two transitions that connect these two levels with the
same $m_{s}=0$ spin sub-level (Raman excitation scheme). The zero-quantum
nature of these transitions is clear by comparing the top trace of
Fig. \ref{Spec}(b) with the middle trace of the same, which was recorded
with a frequency detuning ($\nu_{d}$) of 15 MHz. The transitions
between the $m_{s}=0$ and $m_{s}=\pm1$ manifolds shift correspondingly
by 5 MHz whereas positions of the zero-quantum transitions do not
change. The lower trace of Fig. \ref{Spec}(b) was recorded with the
same pulse sequence of Fig. \ref{Spec}(a) but the flip angle of the
MW pulses is $\pi$ instead of $\pi/2$, which optimizes the signal
from the zero-quantum transitions.

\begin{figure}[t]
\centering \includegraphics[width=8.6cm]{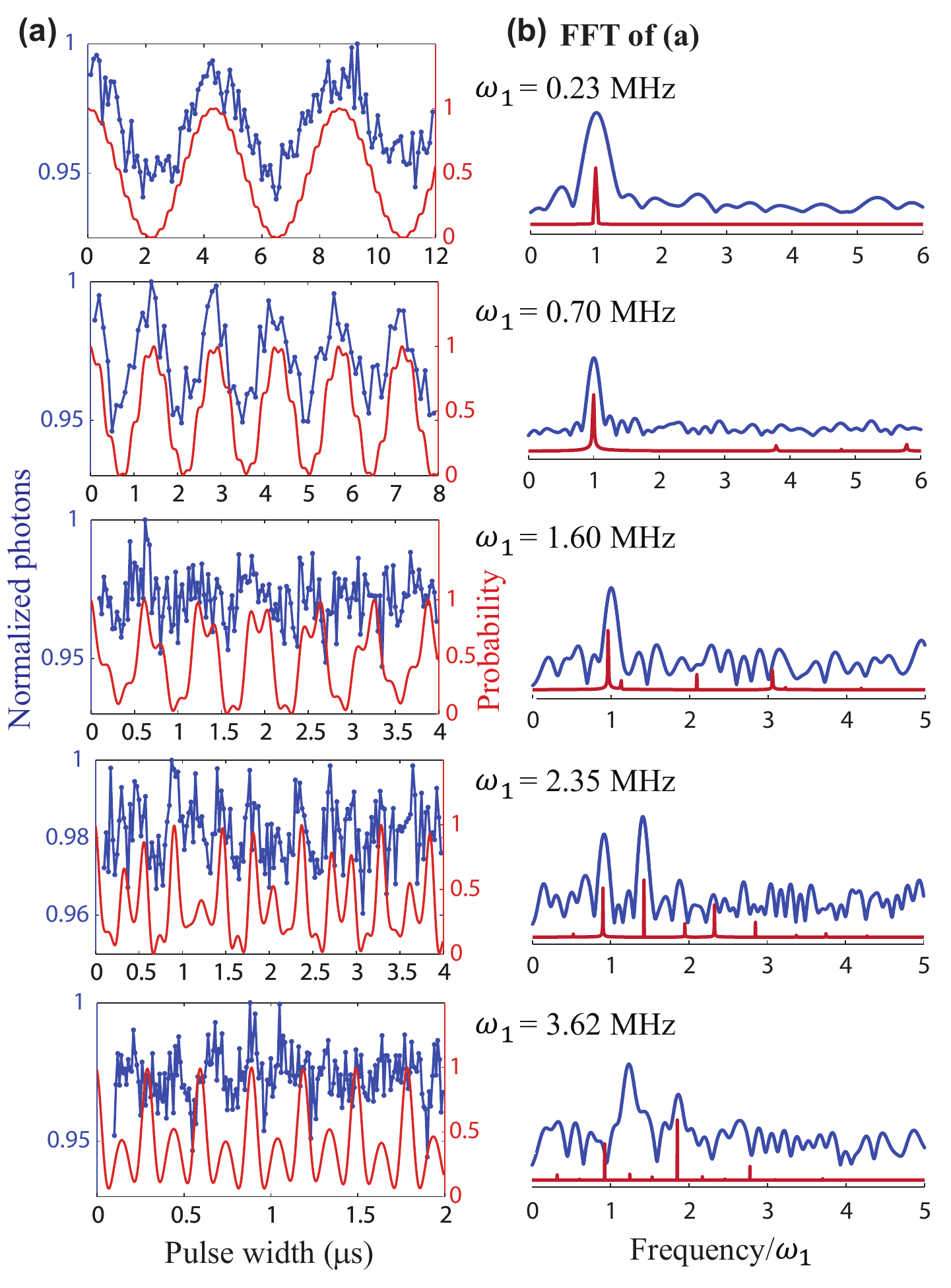} \caption{Dynamics of the two-level system for increasing amplitudes, $\omega_{1}$ of the driving field. (a) Dynamics as a function of the duration of
the RF pulse. The upper blue dots connected by solid lines are the
experimental data and the lower red solid curves are the corresponding
simulated dynamics. (b) Fourier transforms of (a).}
\label{RabiRF} 
\end{figure}

Now, we study the dynamics of the two-level system at the LAC (marked
by solid gray oval in Fig. \ref{Englvl}(c)). These dynamics can be
observed by using the pulse sequence shown in Fig. \ref{pulseq}.
Here, the first laser pulse polarizes the $m_{s}=0$ spin state. The
first MW $\pi$ pulse applied at a frequency of 2873.9 MHz selectively
inverts the transition marked by solid green arrow in Figs. \ref{Englvl}(c)
and \ref{Spec}(b), which is a transition between the $m_{s}=0$ and
$m_{s}=\pm1$ manifolds. This creates a population difference between
the two levels of interest. Then an RF pulse of variable duration
is applied on resonance ($\omega=\omega_{0}=1.7$ MHz) to drive the
transition between the two levels. The corresponding dynamics are
observed by applying another selective MW $\pi$ pulse followed by
the read-out laser pulse, which reads the total population of the
$m_{s}=0$ state. 
The observed dynamics for different amplitudes ($\omega_{1}$) of
the RF pulse are shown in Fig. \ref{RabiRF}(a) (upper blue curves).
The values of $\omega_{1}$ in the strong-driving regime are obtained
by scaling the values measured in the weak-driving regime with the
square root of the RF power. The corresponding dynamics simulated
numerically using Eq. \ref{eq:TLS} are also shown in Fig. \ref{RabiRF}(a)
(lower red curves). The corresponding Fourier transforms are shown
in Fig. \ref{RabiRF}(b). When the amplitude, $\omega_{1}$ of the
driving field is $0.23$ MHz, which is small compared to the transition
frequency, $\omega_{0}$ of the system, the Rabi oscillations of the
system are clearly sinusoidal. As we increase $\omega_{1}$ to 1.6
MHz, which is close to $\omega_{0}$=1.7 MHz, the effects due to the
counter-rotating field component become significant and the oscillations
of the system deviate significantly from the simple sinusoidal behavior.
When $\omega_{1}>\omega_{0}$, the oscillations become strongly anharmonic
as the effects due to the counter-rotating field component become
more prominent. In this strong-driving regime, the system's dynamics
contain frequency components that are significantly higher than the
amplitude $\omega_{1}$ of the field. This anharmonic and non-linear
behavior of the system can be seen from Fig. \ref{RabiRF} when $\omega_{1}=$2.35
and 3.62 MHz: the system dynamics now includes multiple frequencies,
some of which are higher than $\omega_{1}$. The results of the numerical
simulations, which are also shown in Fig. \ref{RabiRF} are in good
agreement with the experimental observations. However, there is some
deviation between the experimental results and the simulated ones,
in particular when $\omega_{1}=3.62$ MHz. This can be explained as
follows.

In the strong-driving regime, the dynamics are very sensitive to the
amplitude, phase and shape of the RF pulse. In the present case, the
deviations between the experiment and the simulation are mainly due
to the effect of the phase of the RF pulse. Each data point of the
experimental time-domain data was obtained by taking the average value
over 200,000 repetitions of the pulse sequence shown in Fig. \ref{pulseq}.
The RF pulse of this pulse sequence was applied by chopping a continuous
RF signal with an RF switch. This implies that the phase of this pulse
is not the same in all the repetitions. This has no significant effect
on the dynamics in the weak-driving regime. However, in the strong-driving
regime, in particular when $\omega_{1}>\omega_{0}$, this has significant
effects. However, the phase of the RF pulse only alters the intensity
pattern of the spectral peaks in the Fourier-transformed dynamics not
their positions (frequencies). This is explained in detail by using
numerical simulations, in Appendix for the case $\omega_{1}=3.62$
MHz. This implies that the frequencies of the identifiable peaks of
the experimental spectrum should match the frequencies of the peaks
in the simulated one for any phase but their amplitudes can be
different. This can be observed in Fig. \ref{RabiRF}. When $\omega_{1}=3.62$
MHz, there is a strong peak in the experimental spectrum, roughly at
frequency/$\omega_{1}=1.2$. At this position, there is also a peak
in the simulated spectrum, but it is of very low intensity for the
phase ($0^{\circ}$) used in the simulation. This experimental spectrum
has a better match with the simulated one averaged over many different phases of the driving field (data shown in the Appendix).

\section{Conclusion}

We studied, theoretically and experimentally, the anharmonic and non-linear dynamics of
a two-level system of a single spin driven in the strong-driving regime,
where the rotating-wave approximation is not valid. Studying spin
dynamics in this regime is interesting not only from the fundamental
physics perspective but also to the quantum information processing
because of the ultra-fast quantum gates possible in this regime. The
two-level system explored here will be useful to experimentally
study strong-driving dynamics for very high ratios of driving field's amplitude to the transition frequency of the system. This system will also be useful as a good test bed to explore time-optimal pulse shapes in the strong-driving regime by using optimal control theory techniques \cite{Khaneja2001PRA,MasonOptCont,CRAB1,CRAB2,BRWAJelezko}.

\section{Acknowledgments}

We acknowledge experimental assistance from Fabian Lehmann for the
construction of the magnetic field rotation setup. We thank Jingfu
Zhang for useful discussions. This work was supported by the DFG through
grant no. Su 192/31-1.

\bibliography{bibNV1,TLS1Notes}

\begin{thebibliography}{27}%
\makeatletter
\providecommand \@ifxundefined [1]{%
 \@ifx{#1\undefined}
}%
\providecommand \@ifnum [1]{%
 \ifnum #1\expandafter \@firstoftwo
 \else \expandafter \@secondoftwo
 \fi
}%
\providecommand \@ifx [1]{%
 \ifx #1\expandafter \@firstoftwo
 \else \expandafter \@secondoftwo
 \fi
}%
\providecommand \natexlab [1]{#1}%
\providecommand \enquote  [1]{``#1''}%
\providecommand \bibnamefont  [1]{#1}%
\providecommand \bibfnamefont [1]{#1}%
\providecommand \citenamefont [1]{#1}%
\providecommand \href@noop [0]{\@secondoftwo}%
\providecommand \href [0]{\begingroup \@sanitize@url \@href}%
\providecommand \@href[1]{\@@startlink{#1}\@@href}%
\providecommand \@@href[1]{\endgroup#1\@@endlink}%
\providecommand \@sanitize@url [0]{\catcode `\\12\catcode `\$12\catcode
  `\&12\catcode `\#12\catcode `\^12\catcode `\_12\catcode `\%12\relax}%
\providecommand \@@startlink[1]{}%
\providecommand \@@endlink[0]{}%
\providecommand \url  [0]{\begingroup\@sanitize@url \@url }%
\providecommand \@url [1]{\endgroup\@href {#1}{\urlprefix }}%
\providecommand \urlprefix  [0]{URL }%
\providecommand \Eprint [0]{\href }%
\providecommand \doibase [0]{http://dx.doi.org/}%
\providecommand \selectlanguage [0]{\@gobble}%
\providecommand \bibinfo  [0]{\@secondoftwo}%
\providecommand \bibfield  [0]{\@secondoftwo}%
\providecommand \translation [1]{[#1]}%
\providecommand \BibitemOpen [0]{}%
\providecommand \bibitemStop [0]{}%
\providecommand \bibitemNoStop [0]{.\EOS\space}%
\providecommand \EOS [0]{\spacefactor3000\relax}%
\providecommand \BibitemShut  [1]{\csname bibitem#1\endcsname}%
\let\auto@bib@innerbib\@empty
\bibitem [{\citenamefont {Slichter}(1990)}]{Slichter}%
  \BibitemOpen
  \bibfield  {author} {\bibinfo {author} {\bibfnamefont {C.~P.}\ \bibnamefont
  {Slichter}},\ }\href@noop {} {\emph {\bibinfo {title} {Principles of Magnetic
  Resonance}}}\ (\bibinfo  {publisher} {Springer},\ \bibinfo {year}
  {1990})\BibitemShut {NoStop}%
\bibitem [{\citenamefont {Allen}\ and\ \citenamefont {Eberly}(1987)}]{AE}%
  \BibitemOpen
  \bibfield  {author} {\bibinfo {author} {\bibfnamefont {L.}~\bibnamefont
  {Allen}}\ and\ \bibinfo {author} {\bibfnamefont {J.~H.}\ \bibnamefont
  {Eberly}},\ }\href@noop {} {\emph {\bibinfo {title} {Optical Resonance and
  Two-Level atoms}}}\ (\bibinfo  {publisher} {Dover publications, New York},\
  \bibinfo {year} {1987})\BibitemShut {NoStop}%
\bibitem [{\citenamefont {Nielsen}\ and\ \citenamefont {Chuang}(2000)}]{NC}%
  \BibitemOpen
  \bibfield  {author} {\bibinfo {author} {\bibfnamefont {M.~A.}\ \bibnamefont
  {Nielsen}}\ and\ \bibinfo {author} {\bibfnamefont {I.~L.}\ \bibnamefont
  {Chuang}},\ }\href@noop {} {\emph {\bibinfo {title} {Quantum Computation and
  Quantum Information}}}\ (\bibinfo  {publisher} {Cambridge university press},\
  \bibinfo {year} {2000})\BibitemShut {NoStop}%
\bibitem [{\citenamefont {Grossmann}\ \emph {et~al.}(1991)\citenamefont
  {Grossmann}, \citenamefont {Dittrich}, \citenamefont {Jung},\ and\
  \citenamefont {H\"anggi}}]{CDT}%
  \BibitemOpen
  \bibfield  {author} {\bibinfo {author} {\bibfnamefont {F.}~\bibnamefont
  {Grossmann}}, \bibinfo {author} {\bibfnamefont {T.}~\bibnamefont {Dittrich}},
  \bibinfo {author} {\bibfnamefont {P.}~\bibnamefont {Jung}}, \ and\ \bibinfo
  {author} {\bibfnamefont {P.}~\bibnamefont {H\"anggi}},\ }\href {\doibase
  10.1103/PhysRevLett.67.516} {\bibfield  {journal} {\bibinfo  {journal} {Phys.
  Rev. Lett.}\ }\textbf {\bibinfo {volume} {67}},\ \bibinfo {pages} {516}
  (\bibinfo {year} {1991})}\BibitemShut {NoStop}%
\bibitem [{\citenamefont {Boyd}(2000)}]{BoydTLS}%
  \BibitemOpen
  \bibfield  {author} {\bibinfo {author} {\bibfnamefont {J.~K.}\ \bibnamefont
  {Boyd}},\ }\href {\doibase http://dx.doi.org/10.1063/1.533345} {\bibfield
  {journal} {\bibinfo  {journal} {Journal of Mathematical Physics}\ }\textbf
  {\bibinfo {volume} {41}},\ \bibinfo {pages} {4330} (\bibinfo {year}
  {2000})}\BibitemShut {NoStop}%
\bibitem [{\citenamefont {Ashhab}\ \emph {et~al.}(2007)\citenamefont {Ashhab},
  \citenamefont {Johansson}, \citenamefont {Zagoskin},\ and\ \citenamefont
  {Nori}}]{NoriCDT}%
  \BibitemOpen
  \bibfield  {author} {\bibinfo {author} {\bibfnamefont {S.}~\bibnamefont
  {Ashhab}}, \bibinfo {author} {\bibfnamefont {J.~R.}\ \bibnamefont
  {Johansson}}, \bibinfo {author} {\bibfnamefont {A.~M.}\ \bibnamefont
  {Zagoskin}}, \ and\ \bibinfo {author} {\bibfnamefont {F.}~\bibnamefont
  {Nori}},\ }\href {\doibase 10.1103/PhysRevA.75.063414} {\bibfield  {journal}
  {\bibinfo  {journal} {Phys. Rev. A}\ }\textbf {\bibinfo {volume} {75}},\
  \bibinfo {pages} {063414} (\bibinfo {year} {2007})}\BibitemShut {NoStop}%
\bibitem [{\citenamefont {Childress}\ and\ \citenamefont
  {McIntyre}(2010)}]{Chil2010PRA}%
  \BibitemOpen
  \bibfield  {author} {\bibinfo {author} {\bibfnamefont {L.}~\bibnamefont
  {Childress}}\ and\ \bibinfo {author} {\bibfnamefont {J.}~\bibnamefont
  {McIntyre}},\ }\href {\doibase 10.1103/PhysRevA.82.033839} {\bibfield
  {journal} {\bibinfo  {journal} {Phys. Rev. A}\ }\textbf {\bibinfo {volume}
  {82}},\ \bibinfo {pages} {033839} (\bibinfo {year} {2010})}\BibitemShut
  {NoStop}%
\bibitem [{\citenamefont {Pomeau}\ \emph {et~al.}(1986)\citenamefont {Pomeau},
  \citenamefont {Dorizzi},\ and\ \citenamefont {Grammaticos}}]{TLSchaosPomeau}%
  \BibitemOpen
  \bibfield  {author} {\bibinfo {author} {\bibfnamefont {Y.}~\bibnamefont
  {Pomeau}}, \bibinfo {author} {\bibfnamefont {B.}~\bibnamefont {Dorizzi}}, \
  and\ \bibinfo {author} {\bibfnamefont {B.}~\bibnamefont {Grammaticos}},\
  }\href {\doibase 10.1103/PhysRevLett.56.681} {\bibfield  {journal} {\bibinfo
  {journal} {Phys. Rev. Lett.}\ }\textbf {\bibinfo {volume} {56}},\ \bibinfo
  {pages} {681} (\bibinfo {year} {1986})}\BibitemShut {NoStop}%
\bibitem [{\citenamefont {Eidson}\ and\ \citenamefont
  {Fox}(1986)}]{TLSchaosEidson}%
  \BibitemOpen
  \bibfield  {author} {\bibinfo {author} {\bibfnamefont {J.}~\bibnamefont
  {Eidson}}\ and\ \bibinfo {author} {\bibfnamefont {R.~F.}\ \bibnamefont
  {Fox}},\ }\href {\doibase 10.1103/PhysRevA.34.3288} {\bibfield  {journal}
  {\bibinfo  {journal} {Phys. Rev. A}\ }\textbf {\bibinfo {volume} {34}},\
  \bibinfo {pages} {3288} (\bibinfo {year} {1986})}\BibitemShut {NoStop}%
\bibitem [{\citenamefont {Fuchs}\ \emph {et~al.}(2009)\citenamefont {Fuchs},
  \citenamefont {Dobrovitski}, \citenamefont {Toyli}, \citenamefont
  {Heremans},\ and\ \citenamefont {Awschalom}}]{Fuchs2009}%
  \BibitemOpen
  \bibfield  {author} {\bibinfo {author} {\bibfnamefont {G.~D.}\ \bibnamefont
  {Fuchs}}, \bibinfo {author} {\bibfnamefont {V.~V.}\ \bibnamefont
  {Dobrovitski}}, \bibinfo {author} {\bibfnamefont {D.~M.}\ \bibnamefont
  {Toyli}}, \bibinfo {author} {\bibfnamefont {F.~J.}\ \bibnamefont {Heremans}},
  \ and\ \bibinfo {author} {\bibfnamefont {D.~D.}\ \bibnamefont {Awschalom}},\
  }\href {\doibase 10.1126/science.1181193} {\bibfield  {journal} {\bibinfo
  {journal} {Science}\ }\textbf {\bibinfo {volume} {326}},\ \bibinfo {pages}
  {1520} (\bibinfo {year} {2009})}\BibitemShut {NoStop}%
\bibitem [{\citenamefont {London}\ \emph {et~al.}(2014)\citenamefont {London},
  \citenamefont {Balasubramanian}, \citenamefont {Naydenov}, \citenamefont
  {McGuinness},\ and\ \citenamefont {Jelezko}}]{BRWAJelezkoPRA}%
  \BibitemOpen
  \bibfield  {author} {\bibinfo {author} {\bibfnamefont {P.}~\bibnamefont
  {London}}, \bibinfo {author} {\bibfnamefont {P.}~\bibnamefont
  {Balasubramanian}}, \bibinfo {author} {\bibfnamefont {B.}~\bibnamefont
  {Naydenov}}, \bibinfo {author} {\bibfnamefont {L.~P.}\ \bibnamefont
  {McGuinness}}, \ and\ \bibinfo {author} {\bibfnamefont {F.}~\bibnamefont
  {Jelezko}},\ }\href {\doibase 10.1103/PhysRevA.90.012302} {\bibfield
  {journal} {\bibinfo  {journal} {Phys. Rev. A}\ }\textbf {\bibinfo {volume}
  {90}},\ \bibinfo {pages} {012302} (\bibinfo {year} {2014})}\BibitemShut
  {NoStop}%
\bibitem [{\citenamefont {Scheuer}\ \emph {et~al.}(2014)\citenamefont
  {Scheuer}, \citenamefont {Kong}, \citenamefont {Said}, \citenamefont {Chen},
  \citenamefont {Kurz}, \citenamefont {Marseglia}, \citenamefont {Du},
  \citenamefont {Hemmer}, \citenamefont {Montangero}, \citenamefont {Calarco},
  \citenamefont {Naydenov},\ and\ \citenamefont {Jelezko}}]{BRWAJelezko}%
  \BibitemOpen
  \bibfield  {author} {\bibinfo {author} {\bibfnamefont {J.}~\bibnamefont
  {Scheuer}}, \bibinfo {author} {\bibfnamefont {X.}~\bibnamefont {Kong}},
  \bibinfo {author} {\bibfnamefont {R.~S.}\ \bibnamefont {Said}}, \bibinfo
  {author} {\bibfnamefont {J.}~\bibnamefont {Chen}}, \bibinfo {author}
  {\bibfnamefont {A.}~\bibnamefont {Kurz}}, \bibinfo {author} {\bibfnamefont
  {L.}~\bibnamefont {Marseglia}}, \bibinfo {author} {\bibfnamefont
  {J.}~\bibnamefont {Du}}, \bibinfo {author} {\bibfnamefont {P.~R.}\
  \bibnamefont {Hemmer}}, \bibinfo {author} {\bibfnamefont {S.}~\bibnamefont
  {Montangero}}, \bibinfo {author} {\bibfnamefont {T.}~\bibnamefont {Calarco}},
  \bibinfo {author} {\bibfnamefont {B.}~\bibnamefont {Naydenov}}, \ and\
  \bibinfo {author} {\bibfnamefont {F.}~\bibnamefont {Jelezko}},\ }\href
  {http://stacks.iop.org/1367-2630/16/i=9/a=093022} {\bibfield  {journal}
  {\bibinfo  {journal} {New Journal of Physics}\ }\textbf {\bibinfo {volume}
  {16}},\ \bibinfo {pages} {093022} (\bibinfo {year} {2014})}\BibitemShut
  {NoStop}%
\bibitem [{\citenamefont {Deng}\ \emph {et~al.}(2015)\citenamefont {Deng},
  \citenamefont {Orgiazzi}, \citenamefont {Shen}, \citenamefont {Ashhab},\ and\
  \citenamefont {Lupascu}}]{BRWASCqubit}%
  \BibitemOpen
  \bibfield  {author} {\bibinfo {author} {\bibfnamefont {C.}~\bibnamefont
  {Deng}}, \bibinfo {author} {\bibfnamefont {J.-L.}\ \bibnamefont {Orgiazzi}},
  \bibinfo {author} {\bibfnamefont {F.}~\bibnamefont {Shen}}, \bibinfo {author}
  {\bibfnamefont {S.}~\bibnamefont {Ashhab}}, \ and\ \bibinfo {author}
  {\bibfnamefont {A.}~\bibnamefont {Lupascu}},\ }\href {\doibase
  10.1103/PhysRevLett.115.133601} {\bibfield  {journal} {\bibinfo  {journal}
  {Phys. Rev. Lett.}\ }\textbf {\bibinfo {volume} {115}},\ \bibinfo {pages}
  {133601} (\bibinfo {year} {2015})}\BibitemShut {NoStop}%
\bibitem [{\citenamefont {Barfuss}\ \emph {et~al.}(2015)\citenamefont
  {Barfuss}, \citenamefont {Teissier}, \citenamefont {Neu}, \citenamefont
  {Nunnenkamp},\ and\ \citenamefont {Maletinsky}}]{MechDriveNV}%
  \BibitemOpen
  \bibfield  {author} {\bibinfo {author} {\bibfnamefont {A.}~\bibnamefont
  {Barfuss}}, \bibinfo {author} {\bibfnamefont {J.}~\bibnamefont {Teissier}},
  \bibinfo {author} {\bibfnamefont {E.}~\bibnamefont {Neu}}, \bibinfo {author}
  {\bibfnamefont {A.}~\bibnamefont {Nunnenkamp}}, \ and\ \bibinfo {author}
  {\bibfnamefont {P.}~\bibnamefont {Maletinsky}},\ }\href {\doibase
  10.1038/nphys3411} {\bibfield  {journal} {\bibinfo  {journal} {Nat Phys}\
  }\textbf {\bibinfo {volume} {11}},\ \bibinfo {pages} {820} (\bibinfo {year}
  {2015})}\BibitemShut {NoStop}%
\bibitem [{\citenamefont {Laucht}\ \emph {et~al.}(2016)\citenamefont {Laucht},
  \citenamefont {Simmons}, \citenamefont {Kalra}, \citenamefont {Tosi},
  \citenamefont {Dehollain}, \citenamefont {Muhonen}, \citenamefont {Freer},
  \citenamefont {Hudson}, \citenamefont {Itoh}, \citenamefont {Jamieson},
  \citenamefont {McCallum}, \citenamefont {Dzurak},\ and\ \citenamefont
  {Morello}}]{BRWASi}%
  \BibitemOpen
  \bibfield  {author} {\bibinfo {author} {\bibfnamefont {A.}~\bibnamefont
  {Laucht}}, \bibinfo {author} {\bibfnamefont {S.}~\bibnamefont {Simmons}},
  \bibinfo {author} {\bibfnamefont {R.}~\bibnamefont {Kalra}}, \bibinfo
  {author} {\bibfnamefont {G.}~\bibnamefont {Tosi}}, \bibinfo {author}
  {\bibfnamefont {J.~P.}\ \bibnamefont {Dehollain}}, \bibinfo {author}
  {\bibfnamefont {J.~T.}\ \bibnamefont {Muhonen}}, \bibinfo {author}
  {\bibfnamefont {S.}~\bibnamefont {Freer}}, \bibinfo {author} {\bibfnamefont
  {F.~E.}\ \bibnamefont {Hudson}}, \bibinfo {author} {\bibfnamefont {K.~M.}\
  \bibnamefont {Itoh}}, \bibinfo {author} {\bibfnamefont {D.~N.}\ \bibnamefont
  {Jamieson}}, \bibinfo {author} {\bibfnamefont {J.~C.}\ \bibnamefont
  {McCallum}}, \bibinfo {author} {\bibfnamefont {A.~S.}\ \bibnamefont
  {Dzurak}}, \ and\ \bibinfo {author} {\bibfnamefont {A.}~\bibnamefont
  {Morello}},\ }\href {\doibase 10.1103/PhysRevB.94.161302} {\bibfield
  {journal} {\bibinfo  {journal} {Phys. Rev. B}\ }\textbf {\bibinfo {volume}
  {94}},\ \bibinfo {pages} {161302} (\bibinfo {year} {2016})}\BibitemShut
  {NoStop}%
\bibitem [{\citenamefont {Shin}\ \emph {et~al.}(2014)\citenamefont {Shin},
  \citenamefont {Butler}, \citenamefont {Wang}, \citenamefont {Avalos},
  \citenamefont {Seltzer}, \citenamefont {Liu}, \citenamefont {Pines},\ and\
  \citenamefont {Bajaj}}]{Bajaj14Nquadrapole}%
  \BibitemOpen
  \bibfield  {author} {\bibinfo {author} {\bibfnamefont {C.~S.}\ \bibnamefont
  {Shin}}, \bibinfo {author} {\bibfnamefont {M.~C.}\ \bibnamefont {Butler}},
  \bibinfo {author} {\bibfnamefont {H.-J.}\ \bibnamefont {Wang}}, \bibinfo
  {author} {\bibfnamefont {C.~E.}\ \bibnamefont {Avalos}}, \bibinfo {author}
  {\bibfnamefont {S.~J.}\ \bibnamefont {Seltzer}}, \bibinfo {author}
  {\bibfnamefont {R.-B.}\ \bibnamefont {Liu}}, \bibinfo {author} {\bibfnamefont
  {A.}~\bibnamefont {Pines}}, \ and\ \bibinfo {author} {\bibfnamefont {V.~S.}\
  \bibnamefont {Bajaj}},\ }\href {\doibase 10.1103/PhysRevB.89.205202}
  {\bibfield  {journal} {\bibinfo  {journal} {Phys. Rev. B}\ }\textbf {\bibinfo
  {volume} {89}},\ \bibinfo {pages} {205202} (\bibinfo {year}
  {2014})}\BibitemShut {NoStop}%
\bibitem [{\citenamefont {Rao}\ and\ \citenamefont
  {Suter}(2016)}]{Koti2016PRB}%
  \BibitemOpen
  \bibfield  {author} {\bibinfo {author} {\bibfnamefont {K.~R.~K.}\
  \bibnamefont {Rao}}\ and\ \bibinfo {author} {\bibfnamefont {D.}~\bibnamefont
  {Suter}},\ }\href {\doibase 10.1103/PhysRevB.94.060101} {\bibfield  {journal}
  {\bibinfo  {journal} {Phys. Rev. B}\ }\textbf {\bibinfo {volume} {94}},\
  \bibinfo {pages} {060101} (\bibinfo {year} {2016})}\BibitemShut {NoStop}%
\bibitem [{\citenamefont {He}\ \emph {et~al.}(1993)\citenamefont {He},
  \citenamefont {Manson},\ and\ \citenamefont {Fisk}}]{Mansion14Nhyperfine}%
  \BibitemOpen
  \bibfield  {author} {\bibinfo {author} {\bibfnamefont {X.-F.}\ \bibnamefont
  {He}}, \bibinfo {author} {\bibfnamefont {N.~B.}\ \bibnamefont {Manson}}, \
  and\ \bibinfo {author} {\bibfnamefont {P.~T.~H.}\ \bibnamefont {Fisk}},\
  }\href {\doibase 10.1103/PhysRevB.47.8816} {\bibfield  {journal} {\bibinfo
  {journal} {Phys. Rev. B}\ }\textbf {\bibinfo {volume} {47}},\ \bibinfo
  {pages} {8816} (\bibinfo {year} {1993})}\BibitemShut {NoStop}%
\bibitem [{\citenamefont {Felton}\ \emph {et~al.}(2009)\citenamefont {Felton},
  \citenamefont {Edmonds}, \citenamefont {Newton}, \citenamefont {Martineau},
  \citenamefont {Fisher}, \citenamefont {Twitchen},\ and\ \citenamefont
  {Baker}}]{Felton2009}%
  \BibitemOpen
  \bibfield  {author} {\bibinfo {author} {\bibfnamefont {S.}~\bibnamefont
  {Felton}}, \bibinfo {author} {\bibfnamefont {A.~M.}\ \bibnamefont {Edmonds}},
  \bibinfo {author} {\bibfnamefont {M.~E.}\ \bibnamefont {Newton}}, \bibinfo
  {author} {\bibfnamefont {P.~M.}\ \bibnamefont {Martineau}}, \bibinfo {author}
  {\bibfnamefont {D.}~\bibnamefont {Fisher}}, \bibinfo {author} {\bibfnamefont
  {D.~J.}\ \bibnamefont {Twitchen}}, \ and\ \bibinfo {author} {\bibfnamefont
  {J.~M.}\ \bibnamefont {Baker}},\ }\href {\doibase 10.1103/PhysRevB.79.075203}
  {\bibfield  {journal} {\bibinfo  {journal} {Phys. Rev. B}\ }\textbf {\bibinfo
  {volume} {79}},\ \bibinfo {pages} {075203} (\bibinfo {year}
  {2009})}\BibitemShut {NoStop}%
\bibitem [{\citenamefont {Chen}\ \emph {et~al.}(2015)\citenamefont {Chen},
  \citenamefont {Hirose},\ and\ \citenamefont {Cappellaro}}]{PaolaPRB2015}%
  \BibitemOpen
  \bibfield  {author} {\bibinfo {author} {\bibfnamefont {M.}~\bibnamefont
  {Chen}}, \bibinfo {author} {\bibfnamefont {M.}~\bibnamefont {Hirose}}, \ and\
  \bibinfo {author} {\bibfnamefont {P.}~\bibnamefont {Cappellaro}},\ }\href
  {\doibase 10.1103/PhysRevB.92.020101} {\bibfield  {journal} {\bibinfo
  {journal} {Phys. Rev. B}\ }\textbf {\bibinfo {volume} {92}},\ \bibinfo
  {pages} {020101} (\bibinfo {year} {2015})}\BibitemShut {NoStop}%
\bibitem [{\citenamefont {Torrey}(1949)}]{Torrey1949}%
  \BibitemOpen
  \bibfield  {author} {\bibinfo {author} {\bibfnamefont {H.~C.}\ \bibnamefont
  {Torrey}},\ }\href {\doibase 10.1103/PhysRev.76.1059} {\bibfield  {journal}
  {\bibinfo  {journal} {Phys. Rev.}\ }\textbf {\bibinfo {volume} {76}},\
  \bibinfo {pages} {1059} (\bibinfo {year} {1949})}\BibitemShut {NoStop}%
\bibitem [{\citenamefont {Bloch}\ and\ \citenamefont
  {Siegert}(1940)}]{B-Sshift1940}%
  \BibitemOpen
  \bibfield  {author} {\bibinfo {author} {\bibfnamefont {F.}~\bibnamefont
  {Bloch}}\ and\ \bibinfo {author} {\bibfnamefont {A.}~\bibnamefont
  {Siegert}},\ }\href {\doibase 10.1103/PhysRev.57.522} {\bibfield  {journal}
  {\bibinfo  {journal} {Phys. Rev.}\ }\textbf {\bibinfo {volume} {57}},\
  \bibinfo {pages} {522} (\bibinfo {year} {1940})}\BibitemShut {NoStop}%
\bibitem [{\citenamefont {Autler}\ and\ \citenamefont
  {Townes}(1955)}]{AutlerTownes}%
  \BibitemOpen
  \bibfield  {author} {\bibinfo {author} {\bibfnamefont {S.~H.}\ \bibnamefont
  {Autler}}\ and\ \bibinfo {author} {\bibfnamefont {C.~H.}\ \bibnamefont
  {Townes}},\ }\href {\doibase 10.1103/PhysRev.100.703} {\bibfield  {journal}
  {\bibinfo  {journal} {Phys. Rev.}\ }\textbf {\bibinfo {volume} {100}},\
  \bibinfo {pages} {703} (\bibinfo {year} {1955})}\BibitemShut {NoStop}%
\bibitem [{\citenamefont {Khaneja}\ \emph {et~al.}(2001)\citenamefont
  {Khaneja}, \citenamefont {Brockett},\ and\ \citenamefont
  {Glaser}}]{Khaneja2001PRA}%
  \BibitemOpen
  \bibfield  {author} {\bibinfo {author} {\bibfnamefont {N.}~\bibnamefont
  {Khaneja}}, \bibinfo {author} {\bibfnamefont {R.}~\bibnamefont {Brockett}}, \
  and\ \bibinfo {author} {\bibfnamefont {S.~J.}\ \bibnamefont {Glaser}},\
  }\href {\doibase 10.1103/PhysRevA.63.032308} {\bibfield  {journal} {\bibinfo
  {journal} {Phys. Rev. A}\ }\textbf {\bibinfo {volume} {63}},\ \bibinfo
  {pages} {032308} (\bibinfo {year} {2001})}\BibitemShut {NoStop}%
\bibitem [{\citenamefont {Boscain}\ and\ \citenamefont
  {Mason}(2006)}]{MasonOptCont}%
  \BibitemOpen
  \bibfield  {author} {\bibinfo {author} {\bibfnamefont {U.}~\bibnamefont
  {Boscain}}\ and\ \bibinfo {author} {\bibfnamefont {P.}~\bibnamefont
  {Mason}},\ }\href {\doibase http://dx.doi.org/10.1063/1.2203236} {\bibfield
  {journal} {\bibinfo  {journal} {Journal of Mathematical Physics}\ }\textbf
  {\bibinfo {volume} {47}},\ \bibinfo {pages} {062101} (\bibinfo {year}
  {2006})}\BibitemShut {NoStop}%
\bibitem [{\citenamefont {Doria}\ \emph {et~al.}(2011)\citenamefont {Doria},
  \citenamefont {Calarco},\ and\ \citenamefont {Montangero}}]{CRAB1}%
  \BibitemOpen
  \bibfield  {author} {\bibinfo {author} {\bibfnamefont {P.}~\bibnamefont
  {Doria}}, \bibinfo {author} {\bibfnamefont {T.}~\bibnamefont {Calarco}}, \
  and\ \bibinfo {author} {\bibfnamefont {S.}~\bibnamefont {Montangero}},\
  }\href {\doibase 10.1103/PhysRevLett.106.190501} {\bibfield  {journal}
  {\bibinfo  {journal} {Phys. Rev. Lett.}\ }\textbf {\bibinfo {volume} {106}},\
  \bibinfo {pages} {190501} (\bibinfo {year} {2011})}\BibitemShut {NoStop}%
\bibitem [{\citenamefont {Caneva}\ \emph {et~al.}(2011)\citenamefont {Caneva},
  \citenamefont {Calarco},\ and\ \citenamefont {Montangero}}]{CRAB2}%
  \BibitemOpen
  \bibfield  {author} {\bibinfo {author} {\bibfnamefont {T.}~\bibnamefont
  {Caneva}}, \bibinfo {author} {\bibfnamefont {T.}~\bibnamefont {Calarco}}, \
  and\ \bibinfo {author} {\bibfnamefont {S.}~\bibnamefont {Montangero}},\
  }\href {\doibase 10.1103/PhysRevA.84.022326} {\bibfield  {journal} {\bibinfo
  {journal} {Phys. Rev. A}\ }\textbf {\bibinfo {volume} {84}},\ \bibinfo
  {pages} {022326} (\bibinfo {year} {2011})}\BibitemShut {NoStop}%
\end{thebibliography}%

\appendix
\section*{Appendix}

\begin{figure}
\includegraphics[width=8cm]{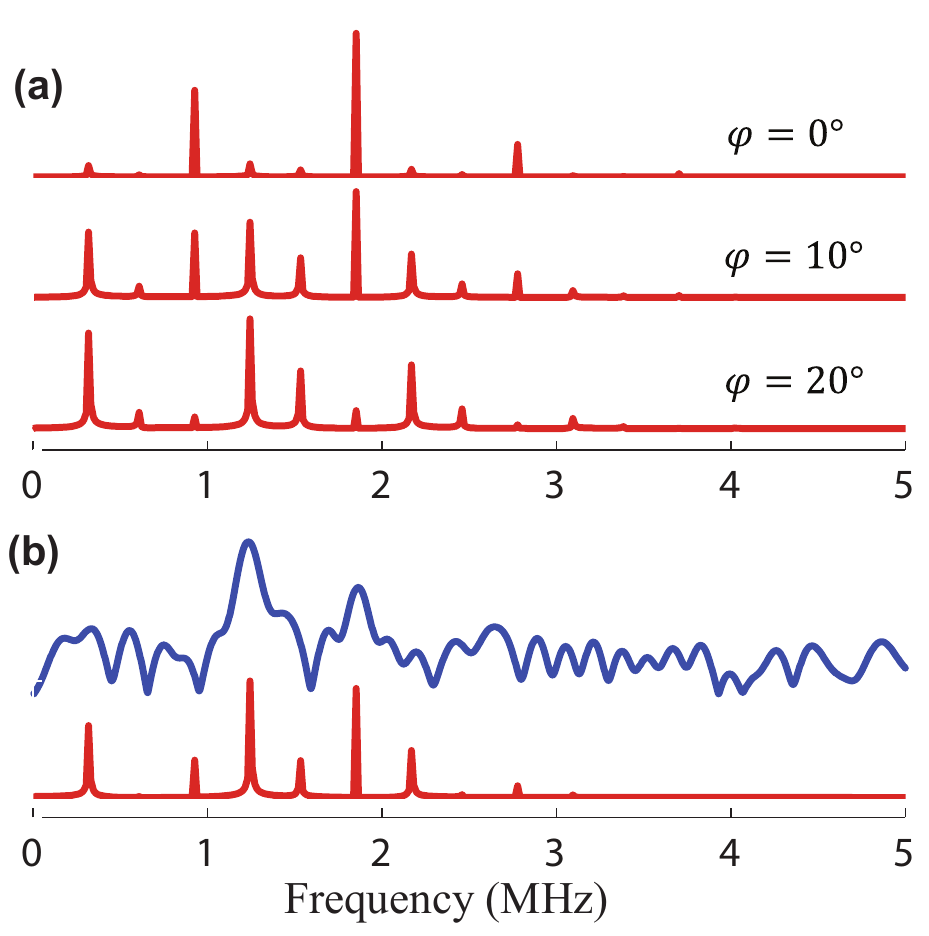}\caption{(a) Fourier transforms of numerically simulated time-domain dynamics
for three different values of $\varphi$ when $\omega_{0}=\omega=1.7$
and $\omega_{1}=3.62$ MHz. (b) Fourier transform of these dynamics
averaged over 100 random values of $\varphi$ between $0$ and $180^{\circ}$
(lower spectrum) in comparison with the experimental one (upper spectrum).}

\label{SpecPhi}
\end{figure}

\textbf{Effect of the phase of the driving field on the dynamics of
the two-level system: }We write the Hamiltonian of the two-level system
as
\[
{\cal H}_{\textrm{TLS}}(t)=\frac{1}{2}\omega_{0}\sigma_{x}+\omega_{1}\cos(2\pi\omega t+\varphi)\sigma_{z}.
\]
This Hamiltonian is the same as in Eq. \ref{eq:TLS} except that the
phase ($\varphi$) of the driving field is included here. In the weak-driving
regime, this phase has no or negligible effect on the dynamics of
the system. However, in the strong-driving regime, in particular when
$\omega_{1}>\omega_{0}$, this phase significantly affects the dynamics.
In order to study this effect, we have numerically simulated the time-domain
dynamics of the above two-level system for different values of $\varphi$
when $\omega_{0}=\omega=1.7$ and $\omega_{1}=3.62$ MHz. The Fourier
transforms of these time-domain dynamics are shown in Fig. \ref{SpecPhi}.
Fig. \ref{SpecPhi}(a) shows the spectra for three different values
of $\varphi$. The intensity patterns of these spectra are significantly
different from one another. However, note that it is only the intensities
of the spectral peaks that change but their positions (frequencies)
remain the same. Fig. \ref{SpecPhi}(b) shows the Fourier transform
of the average time-domain dynamics for 100 random values of $\varphi$
between $0$ and $180^{\circ}$ in comparison with the experimental
one for $\omega_{1}=3.62$ MHz. The match between the experiment and
the simulation is better here than in Fig. \ref{RabiRF}.

\end{document}